\documentstyle[12pt,epsfig]{article}
\newcommand{\SC}{\langle{\cal O}_8^{\psi '}(^1S_0)\rangle}
\newcommand{\SD}{\langle{\cal O}_8^{\psi '}(^3S_1)\rangle}
\newcommand{\PS}{\langle{\cal O}_8^{\psi '}(^3P_0)\rangle}
\newcommand{\SCU}{\langle{\cal O}_8^{\Upsilon '}(^1S_0)\rangle}
\newcommand{\SDU}{\langle{\cal O}_8^{\Upsilon '}(^3S_1)\rangle}
\newcommand{\PSU}{\langle{\cal O}_8^{\Upsilon '}(^3P_0)\rangle}

\input epsf.tex
\def\DESepsf(#1 width #2){\epsfysize=#2 \epsfbox{#1}}
\begin{document}
\begin{flushright}
KOBE--FHD--97--04
December~~~~~~~~1997
\end{flushright}
\begin{center}
{\Large\bf Heavy Quark Productions with Spin\footnote[1]
{Talk given at the International School Seminor'97, 
``Structure of Particles and Nuclei and their Interactions'', 
(Tashkent, Oct. 6--13, 1997)}}  \\
\vspace{1.5em}
{\large\em Toshiyuki Morii} \\
{\small Faculty of Human Development, Kobe University, Nada, 
Kobe 657, Japan}\\
\vspace{1.5em}
{\bf Abstract}
\end{center}
Heavy quark productions in high energy polarized scatterings are
reviewed from a personal point of view.  After mentioning why heavy 
quark physics is so interesting, I concentrate on two rather specific
subjects: (1) polarized $\Lambda_c^+$ productions in deep inelastic
$\ell p$ scatterings and (2) $\psi '$ productions in polarized 
$pp$ collisions.
The first one is an interesting process for extracting the polarized 
gluon in the proton and the second one may give another test of 
the color-octet model of
heavy quarkonium productions in high energy collisions.
\vspace{1.5em}
\begin{flushleft}
{\bf 1. Why heavy quarks?}
\end{flushleft}
Quantum Chromodynamics(QCD) is the underlying field theory of 
strong interactions of quarks and gluons. Although it 
is surprisingly successful in describing physics in perturbative 
regions, it is not so easy task to directly apply QCD for 
nonperturbative regions because of its complicated structure.
To understand physics of quarks and gluons, it is not 
enough to write down
the QCD Lagrangian.  The more important thing is to study how
quarks and gluons interact to make hadrons and how they are produced
in high energy collisions.  

So far, to understand the complicated structure of hadron dymanics in
nonperturbative regions,
many effective theories\cite{Donoghue} 
such as the potential model, the chiral 
perturbation theory, the Skyrme model, the Bag model, the heavy
quark effective theory(HQET) and so on, have been proposed.
These theories have their own scales in which they work.  Here I would like to 
just mention an important role of mass hierarchy and scale.  At present,
we have 6 quark flavors, $u$, $d$, $s$, $c$, $b$, and $t$, whose massses
are arranged, in order, from light to heavy quarks.  
Among them, the masses of $u$, $d$ and
$s$ quarks are smaller than the QCD scale $\Lambda_{QCD}$($\approx 200MeV$)
and the SU(3) flavor symmetry is approximately realized as 
a good symmetry\cite{Georgi}.  Chiral dynamics
works well for these light quarks and can be applied for parameters
remaining to be constant when $m_q\rightarrow 0$.
Actually $m_q$ is not 0 and we have symmetry breaking due to 
$m_q/\Lambda_{QCD}$.  On the other hand, the heavy quark effective 
theory(HQET)\cite{Neubert} works well for the $c$ 
and $b$ quarks which are much heavier
than the $\Lambda_{QCD}$ value.  
For hadrons containing these quarks, 
the heavy quark spin interaction decouples from QCD interactions and thus the 
SU($2N_f$) spin-flavor symmetry appears as a good symmetry.  The HQET can be
applied for parameters remaining to be constant when $m_Q\rightarrow \infty$.
Since $m_Q$ is actually not infinite, we have symmetry breaking due to 
$\Lambda_{QCD}/m_Q$.   Practically, the HQET can not be applied for
hadrons including a top quark.   The top quark is very special 
because it is extremely heavy: it is much heavier than the $W$ boson
with $m_W\approx 80GeV$. 
A top quark decays into a $b$ quark emitting a $W$ boson as a real process
and a single top decay width becomes very large, 
$\Gamma (t\rightarrow bW)\approx 1.5GeV$ for $m_t=175GeV$.  Therefore, 
the life time of the $t$ quark becomes $\approx 10^{-25}sec$ which is 
shorter that the hadronization time $\approx 10^{-23}sec$ and
there is no possibility to make hadrons containing top quarks\cite{Bigi}.

In summary, physics of the $u$, $d$ and $s$ quarks is something 
similar to the solid state physics.  Both of them are described well 
by effective theories with beautiful symmeries.
On the other hand, since a top quark is too heavy to make 
hadrons containing it, we do not need to worry about the complicated 
nonperturbative effect of its hadronization.  
One can directly test the perturbative QCD in top physics.
The situation is, in some sense,
similar to a gas: both of them are rather simple systems.  
However, physics of charm and bottom quarks 
is far from these two limit.  It is something similar to 
physics of liquid which is in between solid and gas.
To understand physics of charm and bottom quarks, we need knowledge
of both perturbative and nonperturbative QCD.  In other word,
heavy quarks, i.e. charm and bottom quarks provide an important playground 
for testing both perturbative and nonperturbative QCD.  
For the nonperturbative and static region, 
the potential model approach is still effective
in the spectroscopy of these heavy flavored hadrons in addition to 
the HQET\cite{Matsuki}.
On the other hand, the heavy quarks are produced only via gluon 
interactions in high energy scatterings and thus play an important role in
extracting an information on perturbative QCD.

Recently, there have been 
new interests in physics of heavy flavored quarks: how they work
effectively for extracting the information on the spin structure 
of nucleons and also how they are produced in high energy collisions.
In this talk, I concentrate on two topics related to these subjects 
which we have recently studied.

\vspace{1.5em}
\begin{flushleft}
{\bf 2. Polarized structure functions of nucleons}
\end{flushleft}
Recent high energy polarized experiments have revealed a much more 
fruitful structure of nucleons than ever considered.  One of the 
most serious problems is so-called the spin puzzle\cite{Cheng}, i.e.
\begin{eqnarray}
\frac{1}{2}&=&\frac{1}{2}\Delta\Sigma+\Delta g
+\langle L_Z\rangle_q+\langle L_Z\rangle_g~,\\
\Delta\Sigma&=&\Delta u+\Delta d+\Delta s\approx 0.3,\\
\Delta s&\approx &-0.12,
\end{eqnarray}
where $\Delta\Sigma$ and $\Delta g$ are the amount of the proton spin carried
by quarks and gluons, respectevely, and
$\langle L_Z\rangle_{q, g}$ implies the orbital angular momenta of quarks
and gluons.
Although there have been many theoretical and 
experimental studies so far, many questions are 
still to be answered: where does the proton spin 
come from?, why are $s$ quarks polarized negatively?, what about gluons?,
how does QCD works? and so on. 

In order to go beyond the present understanding on the nucleon spin 
structure, it is very important to measure the polarized gluon and sea-quark
distribution.  Here I am interested in the gluon polarization in the proton.
Although there have been many discussions on the gluon polarization,
knowledge of the magnitude $\Delta g$ and the behavior 
$\delta g(x, Q^2)$ is still poor. 
Among many processes which are sensitive to
$\delta g$, here we propose a different process(fig.1),

\clearpage
\noindent
\begin{figure}[htb]
  \begin{center}
    \vspace{-0.5cm}
    \hspace{0.1cm}
    \epsfxsize=10.0cm
    \epsfysize=8cm
    \epsfbox{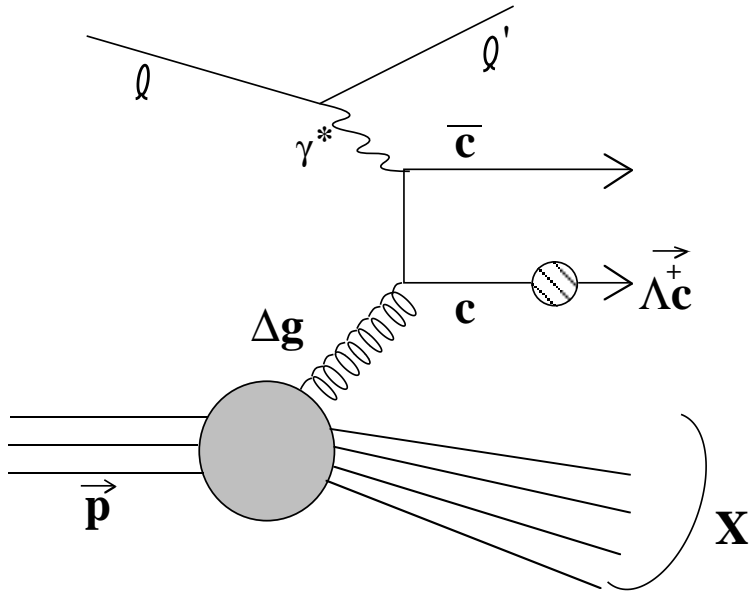}
  \end{center}
  \vspace{-0.6cm}
  \caption{The lowest order QCD diagram for $\Lambda_c^+$ leptoproductions 
in unpolarized lepton-polarized proton scatterings.}

\end{figure}
 
\begin{equation}
\ell +\vec p\rightarrow \vec \Lambda_c^++X,
\end{equation}
to get further information on polarized gluons, 
expecting that the process can be observed in the 
forthcoming COMPASS experiment, where the arrow attached to particles
means that these particles are polarized.
The process is expected to be effective for testing $\delta g$ 
since its cross section is directly proportional to $\delta g$.
Furthermore, the spin of $\Lambda_c^+$ is carried by the $c$ quark
and thus measurement of polarization of $\Lambda_c^+$ in the target
fragmentation region could determine the gluon polarization, $\delta g$.

An interesting parameter is the two-spin asymmetry,
\begin{eqnarray}
A_{LL}~&=&~\frac{\left[d\sigma_{++}-d\sigma_{+-}+
d\sigma_{--}-d\sigma_{-+}\right]}
{\left[d\sigma_{++}+d\sigma_{+-}+
d\sigma_{--}+d\sigma_{-+}\right]}
=~\frac{d\Delta\sigma/dy}{d\sigma/dy}~,
\end{eqnarray}
where $d\sigma_{+-}$, for instance, denotes that the spin
of the target proton and produced $\Lambda_c^+$ is positive and
negative, respectively.
The explicit expressions of the spin-independent and spin-dependent 
cross sections are given in ref.\cite{Kochelev}.
Using the typical examples of polarized gluon 
distributions(GS95\cite{GS95}, BBS95\cite{BBS95}, 
GRV95\cite{GRV95})(fig.2), we have 
calculated the $A_{LL}$ at a CMS energy of the virtual 
photon--proton collision, $\sqrt s=10$GeV
(which corresponds to $\gamma^*$ energy $\nu=56$GeV)
and a momentum transfer squared
$Q^2=10$GeV$^2$(fig.3), whose kinematical region 
can be covered by COMPASS experiments.
We have found that the $A_{LL}$ largely depends on the behavior
of polarized gluons.  Thus we can say that the process might be promising to
test various models of polarized gluons.
\clearpage
\begin{figure}[htb]
\parbox[t]{0.46\textwidth}
{
   \begin{center}
    \vspace{-0.6cm} 
    \epsfxsize=7.5cm
    \epsfysize=6cm
    \epsfbox{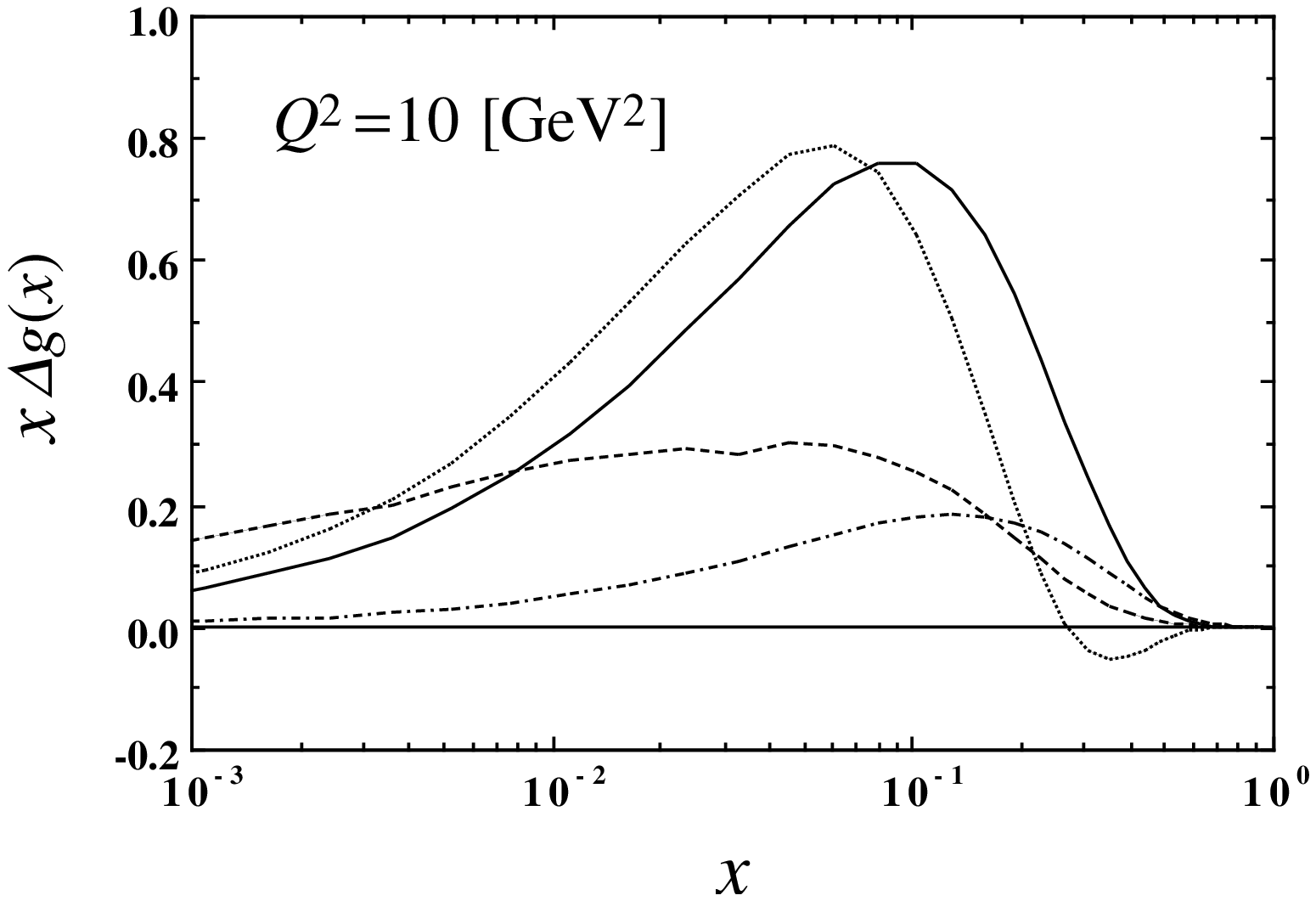}
  \end{center}
  \vspace{-1.1cm}
  \caption{The $x$-dependence of polarized gluon distributions at 
$Q^{2}$=10GeV$^{2}$. The solid, dotted, dash-dotted and dashed lines 
indicate the set A, C of ref.\cite{GS95}, ref.\cite{BBS95} and the 
'standard scenario' of ref.\cite{GRV95}, respectively.}
}
\hspace{0.5cm}
\parbox[t]{0.46\textwidth}
{
  \begin{center}
    \vspace{-0.5cm}
    \epsfxsize=7.5cm
    \epsfysize=6cm
    \epsfbox{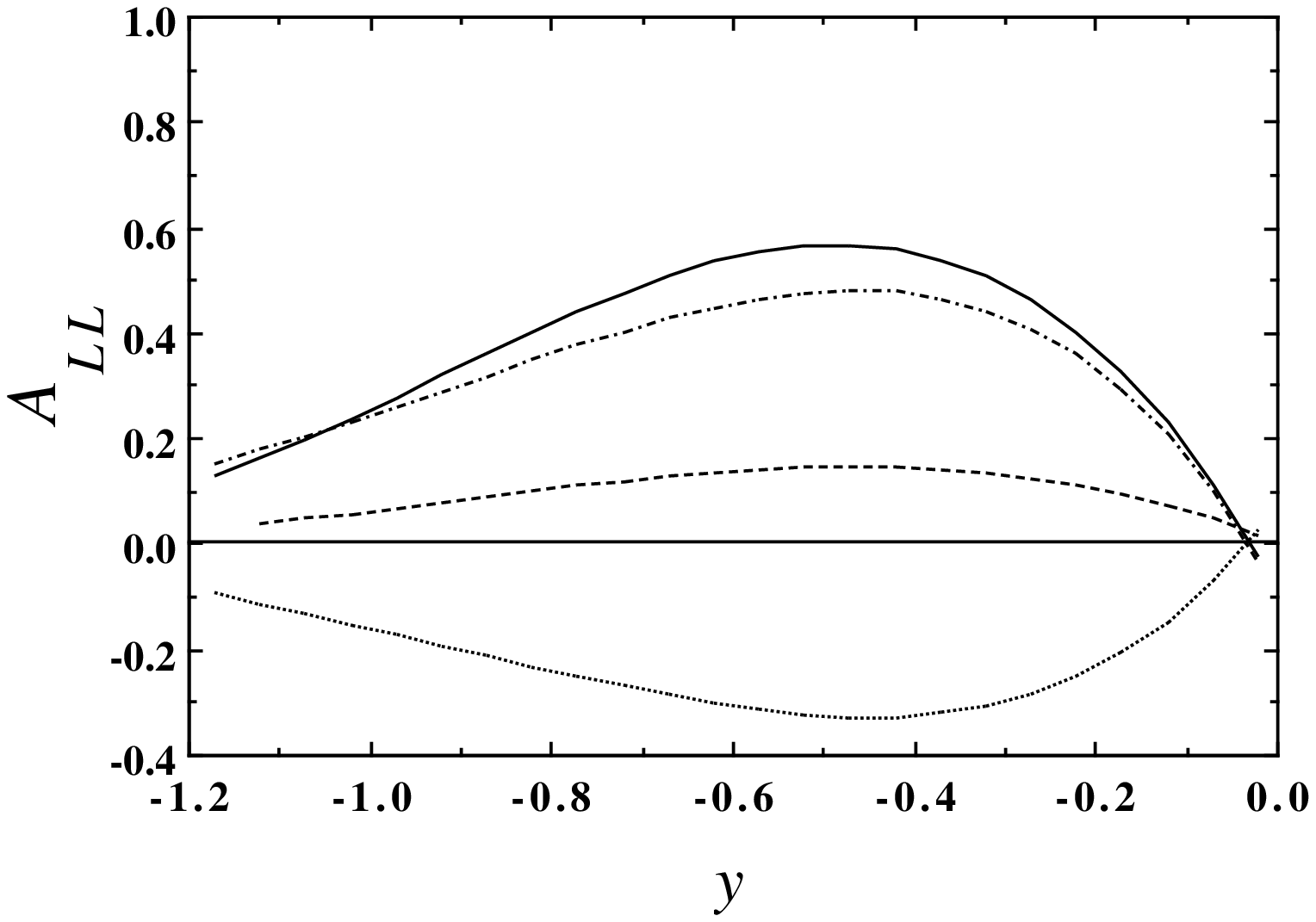}
  \end{center}
  \vspace{-1.2cm}
  \caption{The spin correlation asymmetry $A_{LL}$ as a function 
of rapidty $y$ at $\sqrt{s}$=10GeV and $Q^{2}$=10GeV$^{2}$. Various 
lines represent the same as in fig.2.}
}
\end{figure}
%
\vspace{1.5em}
\begin{flushleft}
{\bf 3. Heavy quarkonium productions}
\end{flushleft}
So far, the standard model is extremely successful in describing the present
experimental data.  However, recent observation\cite{CDF} 
by the CDF collaboration
has shown that the production cross section of charmonium at
large $p_T$ region in $pp$ collisions at Tevatron are order of magnitude
larger than the conventional QCD prediction by the color-singlet model.
This dramatic discrepancy might make a new step toward the deep 
understandings of heavy quarkonium production mechanism.  
To remove such a big discrepancy between the experimental data and the 
prediction of the color--singlet model,   
an interesting new color--octet model has
been proposed recently by several people\cite{octets}.
Physics of the color--octet model whose theoretical ground has been
rigorously formulated by a new effective theory called the 
nonrelativistic QCD(NRQCD)\cite{Bodwin}, 
is now one of the most interesting topics for 
heavy quarkonium productions at high energy.  
Although several processes have been already suggested for testing
the model\cite{Cho}, the discussion seems still controvertial\cite{Cacciari}.
To go beyond the present
theoretical understandings, it is necessary to study other processes. 
To test the model, we propose here another process,
\begin{equation}
\vec p+\vec p\rightarrow \psi '+X, 
\end{equation}
at small $p_T$ regions\cite{Morii}.
Since the process is dominated by the gluon-gluon fusion,
there is no direct production of color-singlet $\psi '$
because of charge conjugation.  
There are only two states which are expected to contribute
to the $\psi '$ production in the final states: the color-octet state
decaying to $\psi ' + g$ and the $2^3P_2$ state decaying to 
$\psi ' + \gamma$.  By using typical examples of 
polarized gluon distributions presented in fig.2, we have 
calculated the two-spin asymmetry $A_{LL}$ for the $\psi '$
produced in this process and found that it is positive 
for the color-octet state and negative for the $2^3P_2$
state.  The spin-dependent and spin-independent cross sections of 
$\psi '$ productions via the color-octet state depend on the parameters, 
$\SC$, $\SD$ and $\PS$\cite{Gupta}, which are 
nonperturbative long--distance factors
associated with the production of a $c\bar c$ 
pair in a color--octet
$^1S_0$, $^3S_1$ and $^3P_0$ states, respectively. 
The model can also be applied likewise even for the $\Upsilon '$ production. 
These nonperturbative factors are obtained 
from recent analysis on charmonium and bottomnium hadroproductions:
\begin{eqnarray}
\SD&\approx &4.6\times 10^{-3}~{\rm [GeV^3]},\\
\SC+\frac{7}{m_c^2}\PS&\approx &5.2\times 10^{-3}~{\rm [GeV^3]},\\ 
\frac{1}{3}\SC+\frac{1}{m_c^2}\PS&\approx &(5.9\pm 1.9)\times
10^{-3}~{\rm [GeV^3]},
\end{eqnarray}
for $\psi '$ productions\cite{Beneke} and 
\begin{eqnarray}
\SDU&\approx &4.1\times 10^{-3}~{\rm [GeV^3]},\\
\SCU+\frac{7}{m_b^2}\PSU&\approx &3.0\times 10^{-2}~{\rm [GeV^3]},\\ 
\frac{1}{5}\SCU+\frac{1}{m_b^2}\PSU&\approx &(9.1\pm 7.2)\times
10^{-3}~{\rm [GeV^3]},
\end{eqnarray}
for $\Upsilon^{\prime}$ productions\cite{Beneke}\cite{Cho2}.
The $A_{LL}$ via the color-octet state largely depend on the 
ratio, $\frac{\tilde\Theta}{\Theta}\equiv
\frac{{\langle{\cal O}_8(^1S_0)\rangle}
-\frac{1}{m^2}{\langle{\cal O}_8(^3P_0)\rangle}}
{{\langle{\cal O}_8(^1S_0)\rangle}+\frac{7}{m^2}
{\langle{\cal O}_8(^3P_0)\rangle}}$, 
whose values range as $\frac{\tilde\Theta}{\Theta}\approx 3.6-8.0$
for $\psi '$ productions and $\frac{\tilde\Theta}{\Theta}\approx -1.88-8.01$ 
for $\Upsilon '$ productions. Those for the 2$^3$P$_2$ state depend on the 
derivative of the 
wave function at the origin, $|R_{2^3P_2}'(0)|$, whose value has been 
estimated by the potential models\cite{Eichten}; 
$|R_{2^3P_2}'(0)|=0.076{\rm GeV^5}-0.186{\rm GeV^5}$ for $\psi '$ and
$|R_{2^3P_2}'(0)|=1.417{\rm GeV^5}-2.067{\rm GeV^5}$ for $\Upsilon^{\prime}$, 
depending on the type of potentials.
We have found that the $A_{LL}$ for the sum of contributions from
the color-octet state and the 2$^3$P$_2$ state becomes
positive for the present parameter regions, in particular,
at $\sqrt s = 50$GeV: the results with typical parameters
are shown in figs.4 and 5.  From this result, one can conclude that 
if we observe a positive $A_{LL}$ in the future RHIC experiment,
we can definitely say that the colo-octet model really contribute 
to this process.  The process is therefore very effective for testing 
the color-octet model.  Since the results largely 
depend on the ratio, 
$\frac{\tilde\Theta}{\Theta}$, it is very important to 
determine the value from other experiments in order to give
a better prediction.
Furthermore, the process is effective for testing
the spin--dependent gluon distribution in the proton 
because its cross section is directly proportional
to the product of $\delta g(x)$ in both protons.

\begin{figure}[htb]
\parbox[t]{0.46\textwidth}
{
   \begin{center}
    \vspace{-0.5cm}
    \epsfxsize=7.5cm
    \epsfysize=6cm
    \epsfbox{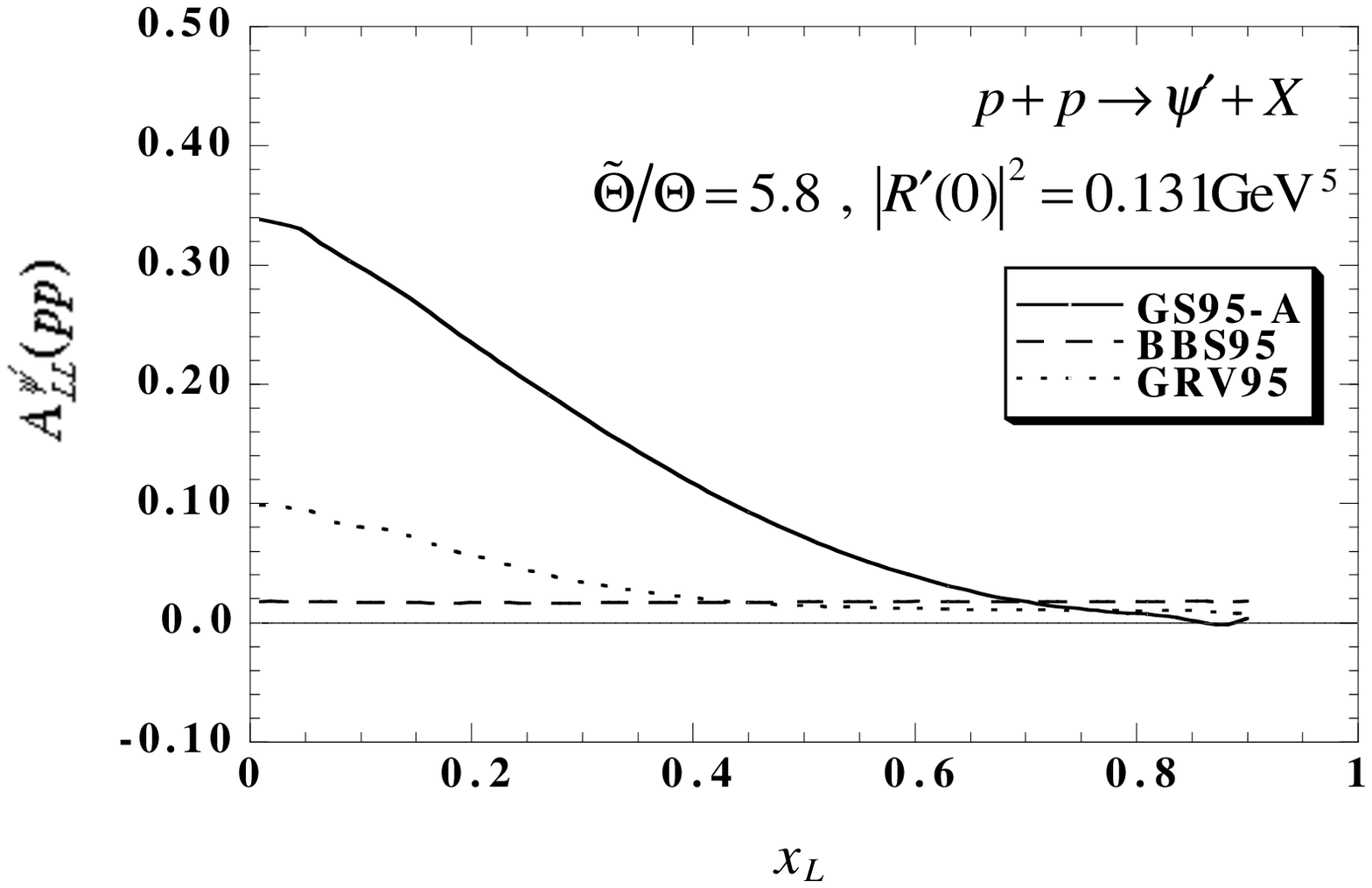}
  \end{center}
  \vspace{-1.0cm}
  \caption{The two--spin asymmetry $A_{LL}^{\psi^{\prime}}(pp)$ 
for $\frac{\tilde\Theta}{\Theta}=5.8$, $|R_{2^3P_2}'(0)|^2=0.131$GeV$^5$ 
at $\sqrt{s}=50$GeV, calculated with various 
types of $\Delta g(x)$, as a function of longitudinal momentum fraction $x_L$ 
of $\psi^{\prime}$. Various lines are indicated in the figure.}
}
\hspace{0.7cm}
\parbox[t]{0.46\textwidth}
{
  \begin{center}
    \vspace{-0.3cm}
    \epsfxsize=7.5cm
    \epsfysize=6cm
    \epsfbox{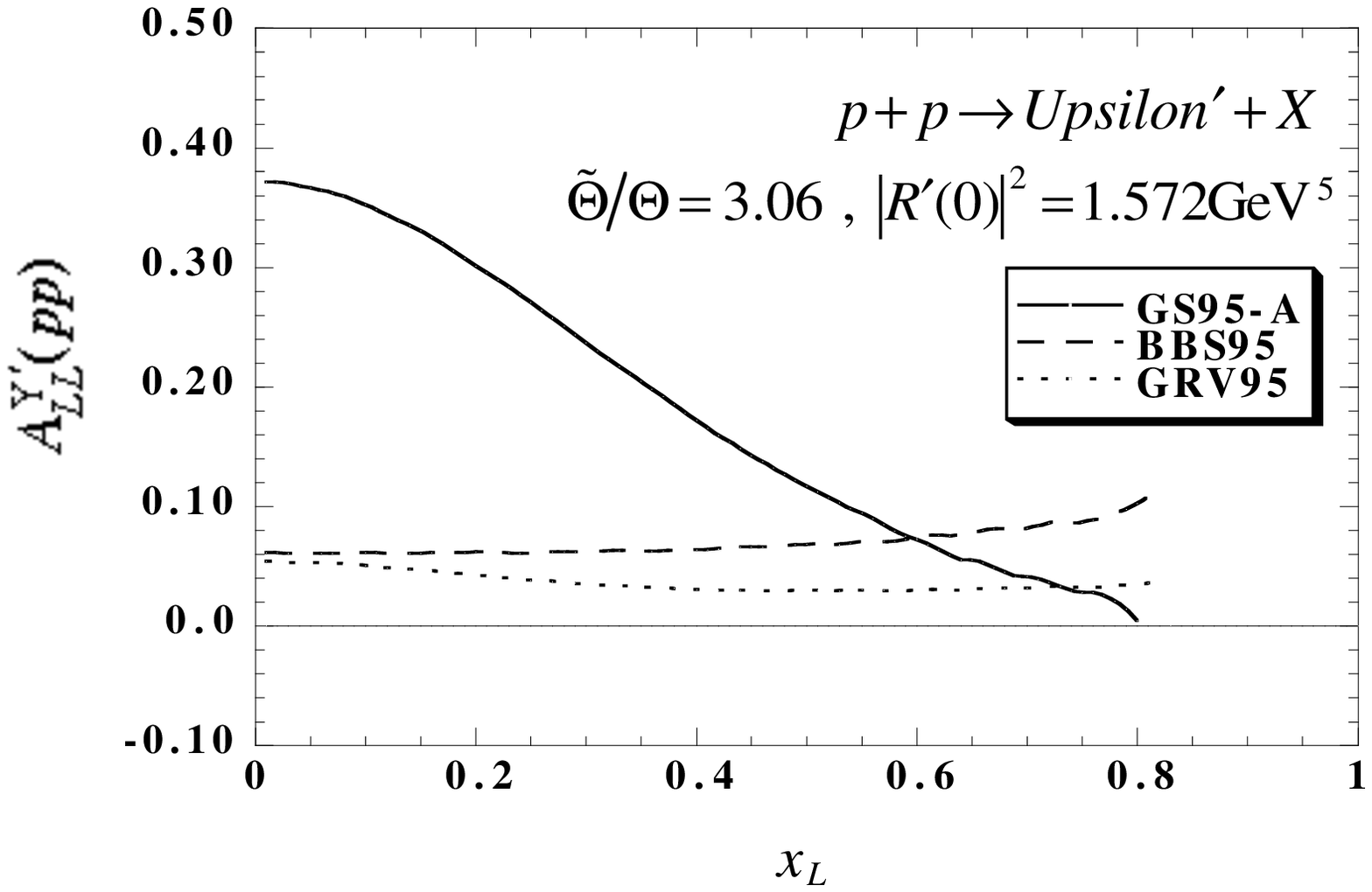}
  \end{center}
  \vspace{-1.2cm}
  \caption{The two--spin asymmetry $A_{LL}^{\Upsilon^{\prime}}(pp)$ 
for $\frac{\tilde\Theta}{\Theta}=3.06$, $|R_{2^3P_2}'(0)|^2=1.572$GeV$^5$ 
at $\sqrt{s}=50$GeV, calculated with various 
types of $\Delta g(x)$, as a function of longitudinal momentum fraction $x_L$ 
of $\Upsilon^{\prime}$. Various lines are indicated in the figure.}
}
\smallskip
\end{figure}

\vspace{1.5em}
\begin{flushleft}
{\bf 4. Outlook}
\end{flushleft}
The production of heavy quarks and quarkonia is a very important subject
to test the standard model and QCD, and furthermore to go beyond the present 
understandings of particle physics.  Polarized experiments are the most 
promising way for testing various theories on heavy quark physics with spin.
In addition to the running experiments, a lot of interesting 
polarized experiments, such as, COMPASS experiment at CERN, 
RHIC project, future polarized HERA experiments, polarized 
$\gamma p$ experiments at SPring-8(Japan), MAMI(Germany), 
TJNAF(USA), and so on, are planned at many places in the world and
will come out soon.  These experiments will lead us to 
a new stage for studying heavy quark physics with spin.

While a lot of experimental and theoretical progress has been done
so far for the spin physics and heavy quark productions, heavy quark
physics is still challenging subject which needs further investigation.

\end{document}